\documentclass[%
reprint,
superscriptaddress,
amsmath,amssymb,
aps,
pra,
floatfix,
]{revtex4-2}

\usepackage{graphicx}
\usepackage{dcolumn}
\usepackage{bm}
\begin{document}

\preprint{APS/123-QED}

\title{Robust mutual synchronization in long spin Hall nano-oscillator chains}

\author{Akash Kumar}
\email{akash.kumar@physics.gu.se}
\affiliation{Physics Department, University of Gothenburg, 412 96 Gothenburg, Sweden.}
\author{Himanshu Fulara}
\affiliation{Department of Physics, Indian Institute of Technology Roorkee, Roorkee 247667, India}
\author{Roman Khymyn}
\affiliation{Physics Department, University of Gothenburg, 412 96 Gothenburg, Sweden.}
\author{Mohammad Zahedinejad}
\affiliation{NanOsc AB, Kista, Sweden.}
\author{Mona Rajabali}
\affiliation{NanOsc AB, Kista, Sweden.}
\author{Xiaotian Zhao}
\affiliation{Physics Department, University of Gothenburg, 412 96 Gothenburg, Sweden.}
\author{Nilamani Behera}
\affiliation{Physics Department, University of Gothenburg, 412 96 Gothenburg, Sweden.}
\author{Afshin Houshang}
\affiliation{Physics Department, University of Gothenburg, 412 96 Gothenburg, Sweden.}
\author{Ahmad A. Awad}
\affiliation{Physics Department, University of Gothenburg, 412 96 Gothenburg, Sweden.}
\author{Johan \AA kerman}
\email{johan.akerman@physics.gu.se}
\affiliation{Physics Department, University of Gothenburg, 412 96 Gothenburg, Sweden.}

\date{\today}

\begin{abstract}
Mutual synchronization of $N$ serially connected spintronic nano-oscillators increases their coherence by a factor $N$ and their output power by $N^2$. Increasing the number of mutually synchronized nano-oscillators in chains is hence of great importance for better signal quality and also for emerging applications such as oscillator-based neuromorphic computing and Ising machines where larger $N$ can tackle larger problems. Here we fabricate spin Hall nano-oscillator chains of up to 50 serially connected nano-constrictions in W/NiFe, W/CoFeB/MgO, and NiFe/Pt stacks and demonstrate robust and complete mutual synchronization of up to 21 nano-constrictions, reaching linewidths of below 200 kHz and quality factors beyond 79,000, while operating at 10 GHz. We also find a square increase in the peak power with the increasing number of mutually synchronized oscillators, resulting in a factor of 400 higher peak power in long chains compared to individual nano-constrictions. Although chains longer than 21 nano-constrictions also show complete mutual synchronization, it is not as robust and their signal quality does not improve as much as they prefer to break up into partially synchronized states. The low current and low field operation of these oscillators along with their wide frequency tunability (2-28 GHz) with both current and magnetic fields, make them ideal candidates for on-chip GHz-range applications and neuromorphic computing.
\end{abstract}

\maketitle

\section{\label{sec:Introduction}Introduction}

Since the advent of spin transfer torque driven magnetization precession in metallic spin valves~\cite{slonczewski1996current,berger1996emission,tsoi2000generation,kiselev2003microwave}, the interest in spintronic microwave oscillators has steadily increased~\cite{chen2016ieeeproc}. 
Mutual synchronization of these non-linear microwave oscillators is of utmost importance for various applications such as efficient ultra-broadband signal generators~\cite{tsunegi2018scaling}, wireless communication, ultra-fast microwave spectral analysis~\cite{litvinenko2020ultrafast,litvinenko2022ultrafast} and recently developed interest in neuromorphic computing among others~\cite{torrejon2017neuromorphic,Romera2018nt}. Moreover, researchers have recently demonstrated energy harvesting from wireless signals using synchronized oscillators in series~\cite{sharma2021electrically}. There have been many attempts to synchronize these oscillators over short and long ranges~\cite{erokhin2014robust,houshang2016spin,lebrun2017mutual,Romera2018nt,tsunegi2018scaling}. However, the complex fabrication process of spin torque nano-oscillators (STNOs) raises a technological issue to scale their synchronization for high frequency applications and hence the progress of synchronizing many STNOs has been rather slow~\cite{lebrun2017mutual,tsunegi2018scaling,sharma2021electrically}.

\begin{figure} [t!]
\centering
\includegraphics[width=\linewidth]{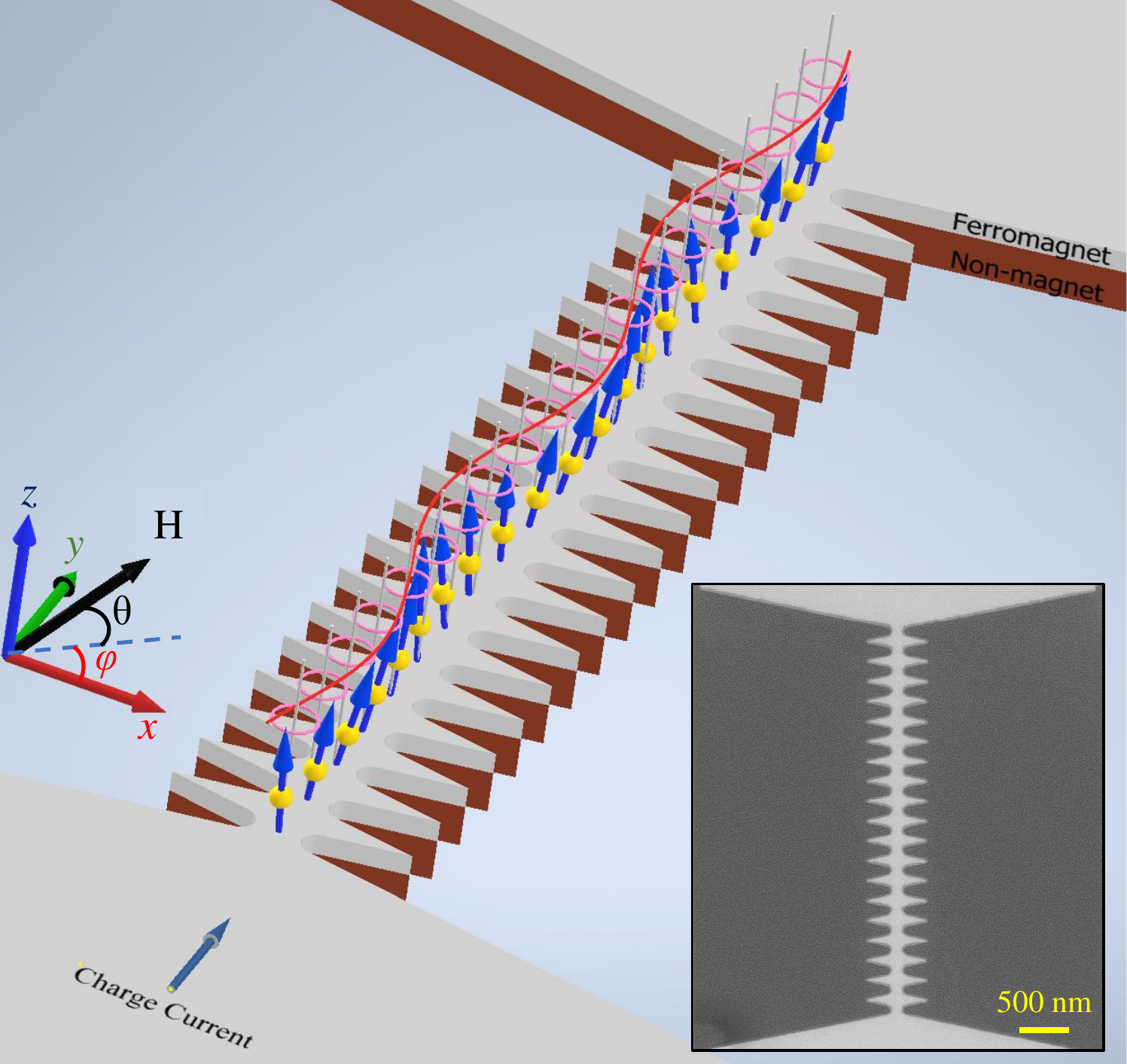}
\caption{\label{fig:Schematic} Schematic representation of 21 nano-constriction SHNOs in a chain fabricated from a non-magnet/ferromagnet bilayer. Blue arrows with yellow spheres represent the local magnetization and its magneto-dynamical precession in each nano-constriction. The charge current is applied in-plane and the magnetic field at an oblique angle. The inset shows a scanning electron microscope image of an actual SHNO chain. }
\end{figure}

\begin{figure*} [t!]
\centering
\includegraphics[width=17cm]{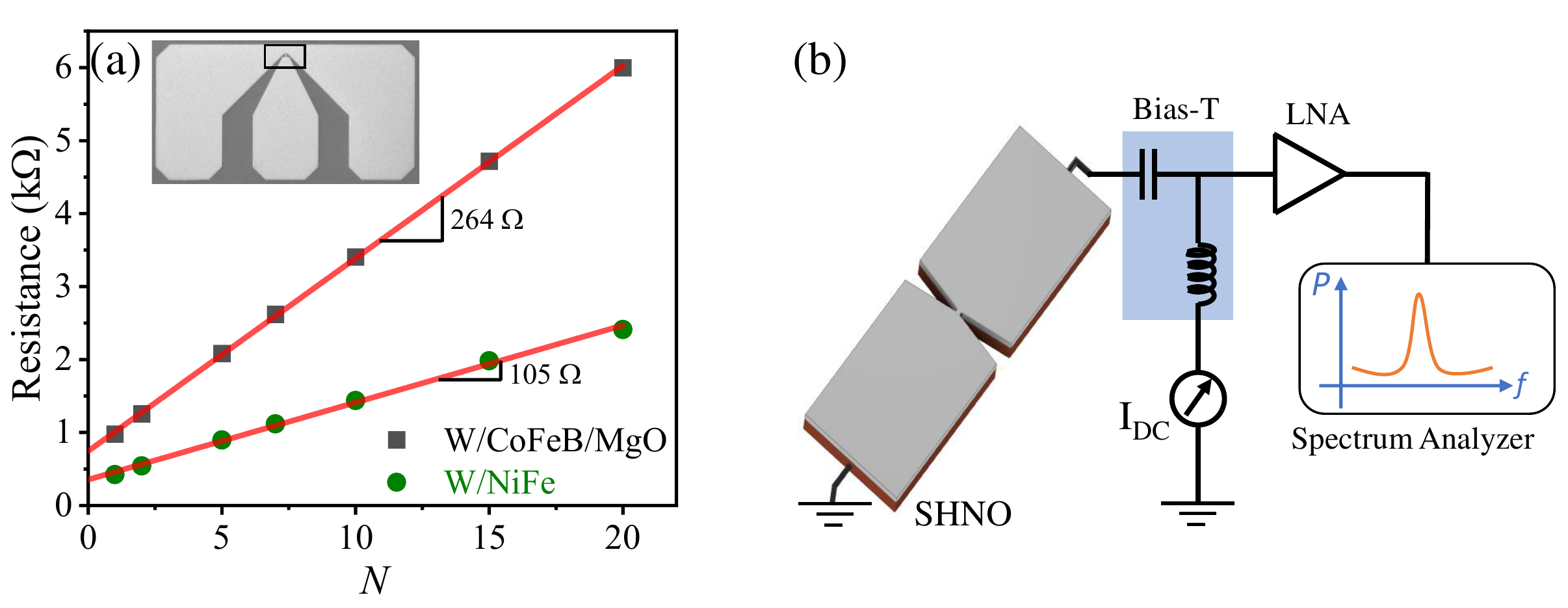}
\caption{\label{fig:Resistance} (a) Resistance of the SHNO chains \emph{vs.}~number of nano-constrictions (\textit{N}), showing the expected linear dependence, with each W/NiFe and W/CoFeB/MgO nano-constriction adding 105 $\Omega$ and 264 $\Omega$, respectively. The inset shows an SEM image of the Ground-Signal-Ground pads connecting to the SHNO chain. (b) Schematic representation of the auto-oscillation measurement set-up. The drive current is provided via a bias-T, which also picks up the generated microwave voltage and feeds it to a low-noise amplifier before being recorded by a spectrum analyzer.}
\end{figure*}

Thanks to the spin Hall effect~\cite{Tserkvovyak2002,Hirsch1999,sinova2015rmp}, a new class of spintronic oscillators, known as spin Hall nano-oscillators (SHNOs), has emerged~\cite{VEDemidov2012,demidov2014nanoconstriction,duan2014nanowire,durrenfeld2017nanoscale}. Compared to STNOs, they rely on the current flowing in-plane, which makes their fabrication easier and allows for a much larger number of SHNOs to synchronize. In particular, nano-constriction based SHNOs~\cite{VEDemidov2012,demidov2014nanoconstriction} can be easily fabricated into 1D chains~\cite{awad2016natphys} and 2D arrays~\cite{zahedinejad2020two}. Earlier work has shown that up to nine SHNOs, separated by 300 nm, can be mutually synchronized to generate both higher output power and a narrower linewidth~\cite{awad2016natphys}. Similarly, 2D arrays of up to 8$\times$8 oscillators~\cite{zahedinejad2020two} were found to synchronize completely. The number of synchronized oscillators along a dimension is hence limited to single digits.

In this work, we study mutual synchronization in much longer SHNO chains of up to 50 serially connected nano-constrictions fabricated from W(5 mm)/CoFeB(1.4 nm)/MgO(2 nm)~\cite{fulara2019spin,Zahedinejad2022natmat,kumar2022fabrication,fulara2020giant}, W(5 nm)/NiFe(3 nm)~\cite{mazraati2016low}, and NiFe(5 nm)/Pt(5 nm)~\cite{awad2016natphys,zahedinejad2020two} material stacks (the order represents the actual stack sequence), focusing primarily on the W based SHNOs with their much lower threshold current. We find that robust and complete mutual synchronization can persist in chains of up to 21 oscillators, resulting in a significantly lower linewidth and higher output power compared to single SHNOs. We also observe mutual synchronization in the longer chains but with deteriorated parameters, which we find to originate from a tendency for the longer chains to separate into shorter mutually synchronized sections.

\section{Results and Discussion}
\subsection{SHNO chains and measurement set-up}
Figure~\ref{fig:Schematic} shows the layout for a chain of nano-constriction SHNOs made from a non-magnet (NM) -- ferromagnet (FM) bilayer. The inset shows a scanning electron micrograph of an actual SHNO chain. A charge current flows in the film plane and a magnetic field is applied at an oblique OOP angle, $\theta$. The spin Hall effect of the NM layer converts the charge current into a transverse spin current exerting an anti-damping torque on the FM layer, which above a certain threshold current can generate auto-oscillations of the local magnetization in each nano-constriction. In this work, we explore 150 nm wide nano-constrictions separated by 200 nm center-to-center separation and chain lengths of up to 50 nano-constrictions. We primarily study and compare chains made from W/CoFeB/MgO and W/NiFe material stacks, where W was chosen for its very large spin-Hall angle ($\theta_{SH}$ = -0.44, for details see supplementary file) and the FM layers for their low damping of $\alpha_{CoFeB}$ = 0.025 and $\alpha_{NiFe}$ = 0.032. We also compare our results with 21 synchronized nano-constrictions in the widely studied NiFe/Pt~\cite{awad2016natphys,zahedinejad2020two} system (shown in the supplementary file), where a much larger charge current density is required because of the lower spin Hall angle of Pt thin films. Figure~\ref{fig:Resistance}(a) shows how the resistance of the SHNO chains increases linearly with the number of nano-constrictions, each nano-constriction adding 105 ohm and 264 ohm of resistance for W/NiFe and W/CoFeB/MgO, respectively. A schematic of the measurement set-up is shown in Fig.~\ref{fig:Resistance}(b). Further details can be found in the Methods section. 

\begin{figure*} [t!]
\centering
\includegraphics[width=17.6cm]{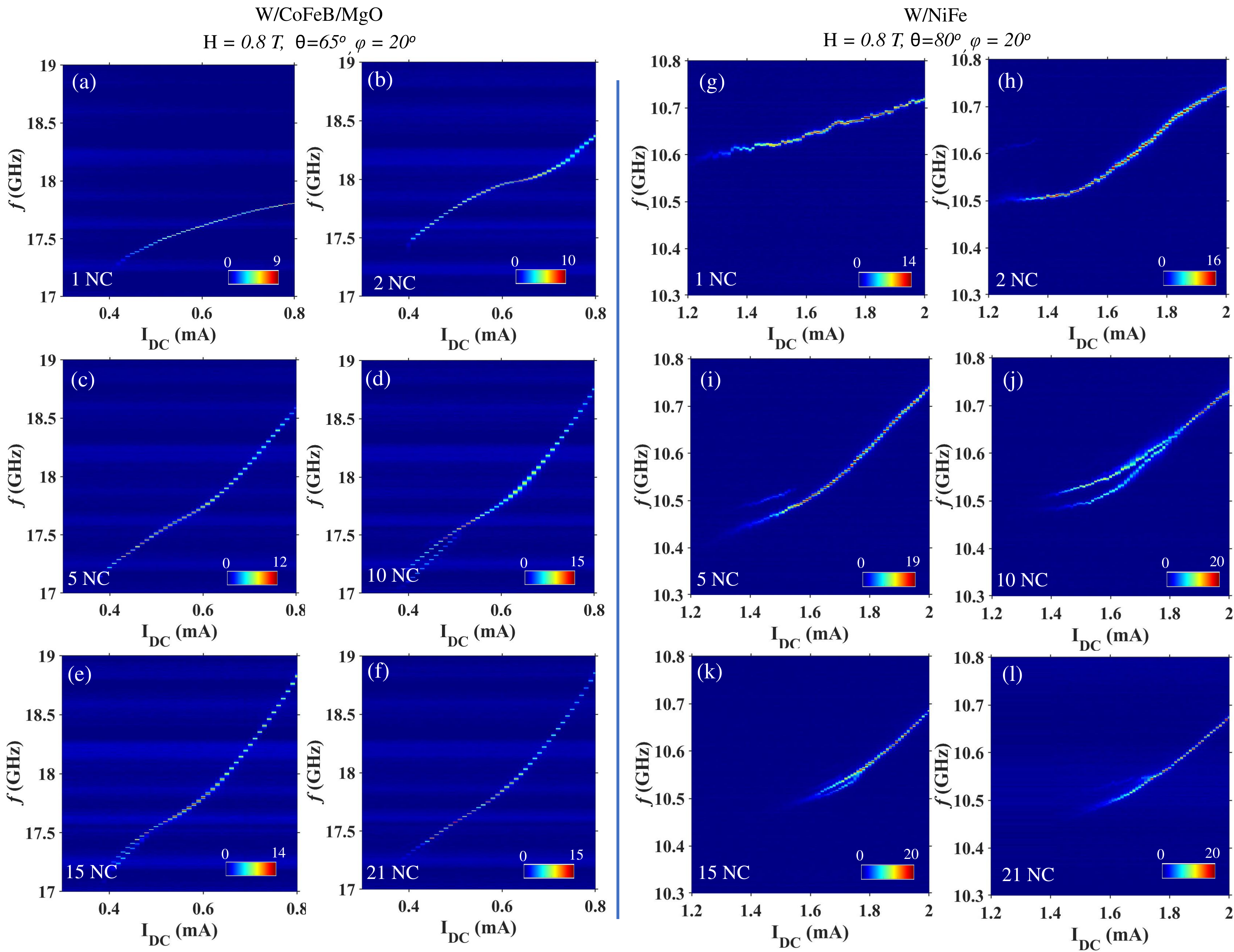}
\caption{\label{fig:PSD} Power spectral density (PSD) of the microwave signal generated by a single nano-constriction and up to 21 mutually synchronized nano-constriction SHNOs for (a-f) W/CoFeB/MgO and (g-l) W/NiFe, respectively. }
\end{figure*}

\subsection{Mutual Synchronization for $N=$ 1--21}

Figure~\ref{fig:PSD} shows the power spectral density (PSD) of auto-oscillation in W/CoFeB/MgO (a-f) and W/NiFe (g-l) based SHNOs with $N=$ 1--21. The behavior of chains with $N\geq$ 30 will be discussed in a later section below. Both types of SHNOs show a positive non-linearity, at given applied field magnitude and angles. As discussed in Ref.~\cite{fulara2019spin}, the perpendicular magnetic anisotropy of the W/CoFeB/MgO stack increases its non-linearity compared to W/NiFe. The non-linearity is also substantially higher for $N \geq$ 5 than for the single and double nano-constrictions, which indicates spin waves emission out of the nano-constriction regions. Thus, there are relatively substantial energy losses for the low number of oscillators by a spin wave emission, which limits the nonlinear frequency shift. Meanwhile, for the larger $N$ the emitted waves contribute energy to the neighboring oscillators, increasing the non-linearity. The increased non-linearity is, at higher currents, accompanied by a change from a concave to a convex current dependence. 

All chains show complete synchronization towards higher currents. The maximum peak power (indicated by the dB over noise scale) also increases with $N$. At lower currents, partial synchronization into primarily two separate signals can be clearly observed. It is noteworthy that the fully synchronized state coincides roughly with the change in curvature. As the change in curvature is only observed in chains and becomes more pronounced for larger $N$, this suggests that, at higher currents, the auto-oscillation mode changes character due to the chain geometry. The robust mutual synchronization is governed by a combination of dipolar and spin wave-mediated coupling. The role of propagating spin waves and their comparison to dipolar coupling was first studied theoretically in nano-contact STNOs~\cite{slavin2009nonlinear}, where analytical calculations suggested a dominant role of propagating spin waves at separations larger than 100 nm. Spin wave beams were also responsible for robust synchronization of up to five nano-contact STNOs \cite{houshang2016spin}. Similar to nano-contact STNOs, nano-constriction SHNOs share a common ferromagnetic layer, which suggests that the same arguments should apply. Following Ref.~\cite{slavin2009nonlinear}, the spin wave coupling strength should then be about twice that of dipolar coupling at a 200 nm separation. 

\begin{figure*} [t!]
\centering
\includegraphics[width=14cm]{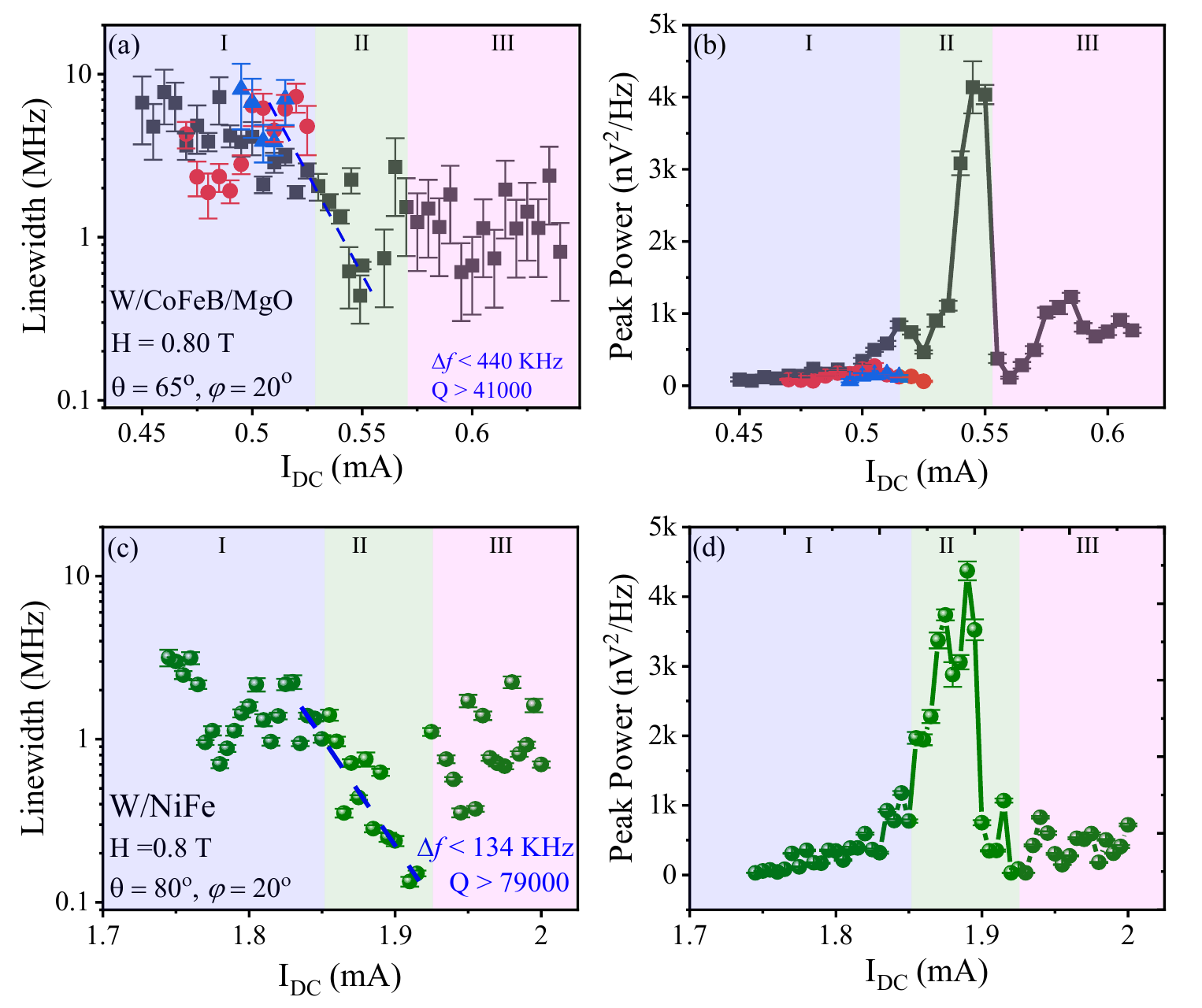}
\caption{\label{fig:21NCs} (a) and (b) Linewidth \emph{vs.}~ I$_{DC}$ and peak power \emph{vs.}~ I$_{DC}$ for 21 mutually synchronized W/CoFeB/MgO SHNOs. (c) and (d) Linewidth \emph{vs.}~ I$_{DC}$ and peak power \emph{vs.}~ I$_{DC}$ for 21 mutually synchronized W/NiFe SHNOs. The dashed blue line represents a complete mutual synchronization of 21 oscillators. The three color-coded regions represent the un-synchronized state (Blue), robustly synchronized state (Olive) and high current unstable synchronized state (pink). }
\end{figure*}

\subsection{Analysis of linewidth and output power}

The microwave signal is fitted with a single Lorentzian function to extract the power and linewidth. Figure~\ref{fig:21NCs} summarizes and compares the linewidth and peak power of 21 mutually synchronized oscillators for W/CoFeB/MgO and W/NiFe thin films, respectively; results for NiFe/Pt thin films are shown in the supplementary. We observe the lowest linewidth of 134 kHz in the synchronized state of the W/NiFe based oscillator [Fig.~\ref{fig:21NCs}(c)], with a peak power just above 4000 nV$^2$/Hz [Fig.~\ref{fig:21NCs}(d)]. The lowest linewidth of the W/CoFeB/MgO chain is 440 kHz [see Fig.~\ref{fig:21NCs}(c)], also with peak power just above 4000 nV$^2$/Hz [Fig.~\ref{fig:21NCs}(d)]. For NiFe/Pt based SHNOs, we observe the lowest linewidth of 275 kHz with enormous peak power of 40,000 nV$^2$/Hz (see supplementary file).

However, once this highest-quality signal is achieved, increasing the current further deteriorates the signal quality to intermediate values. This deterioration does not seem to be related to a loss of mutual synchronization as we only observe a single signal in all devices in this current range. Instead, this behavior coincides with the change in curvature described above, indicating that it could be due to a change in the auto-oscillating mode character. We hence define three different regions: I) incomplete partial synchronization with relatively poor signal quality, II) complete mutual synchronization with the best signal quality, and III) a possible different auto-oscillating regime with intermediate signal quality. 

\begin{figure} [t!]
\centering
\includegraphics[width=8cm]{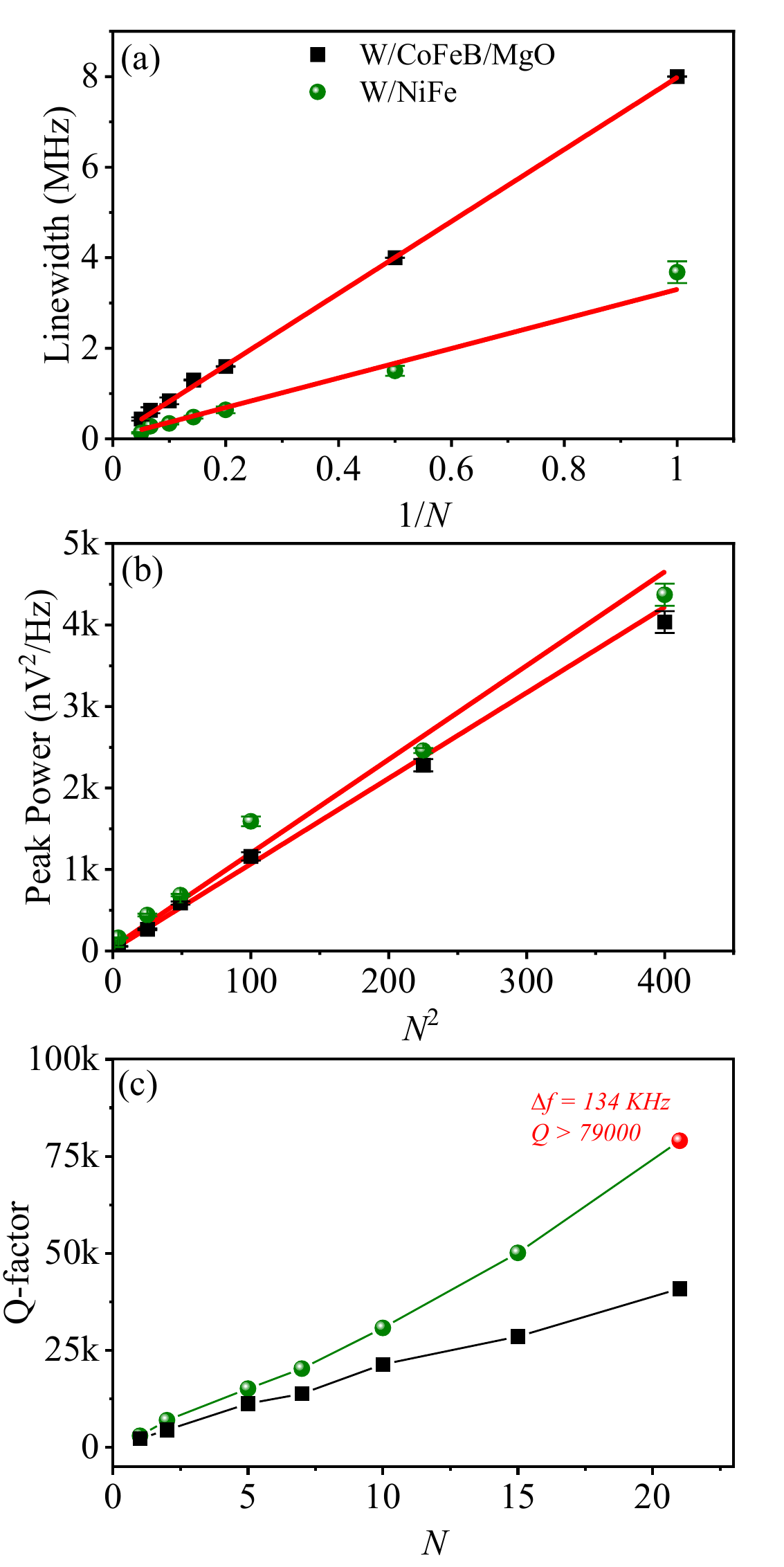}
\caption{\label{fig:parameters} (a) Linewidth \emph{vs.}~inverse of number of nano-constrictions (1/$N$) in a chain of SHNOs for W/CoFeB/MgO and W/NiFe based oscillators. (b) Peak power \emph{vs.}~square of number of oscillators ($N^2$) for W/CoFeB/MgO and W/NiFe based oscillators, respectively. (C) $Q-$factor of SHNOs with number of oscillators ($N$), Black squares are for W/CoFeB/MgO and green circles represents W/NiFe. The red solid line represents linear fit. }
\end{figure}

\begin{figure*} [t!]
\centering
\includegraphics[width=12cm]{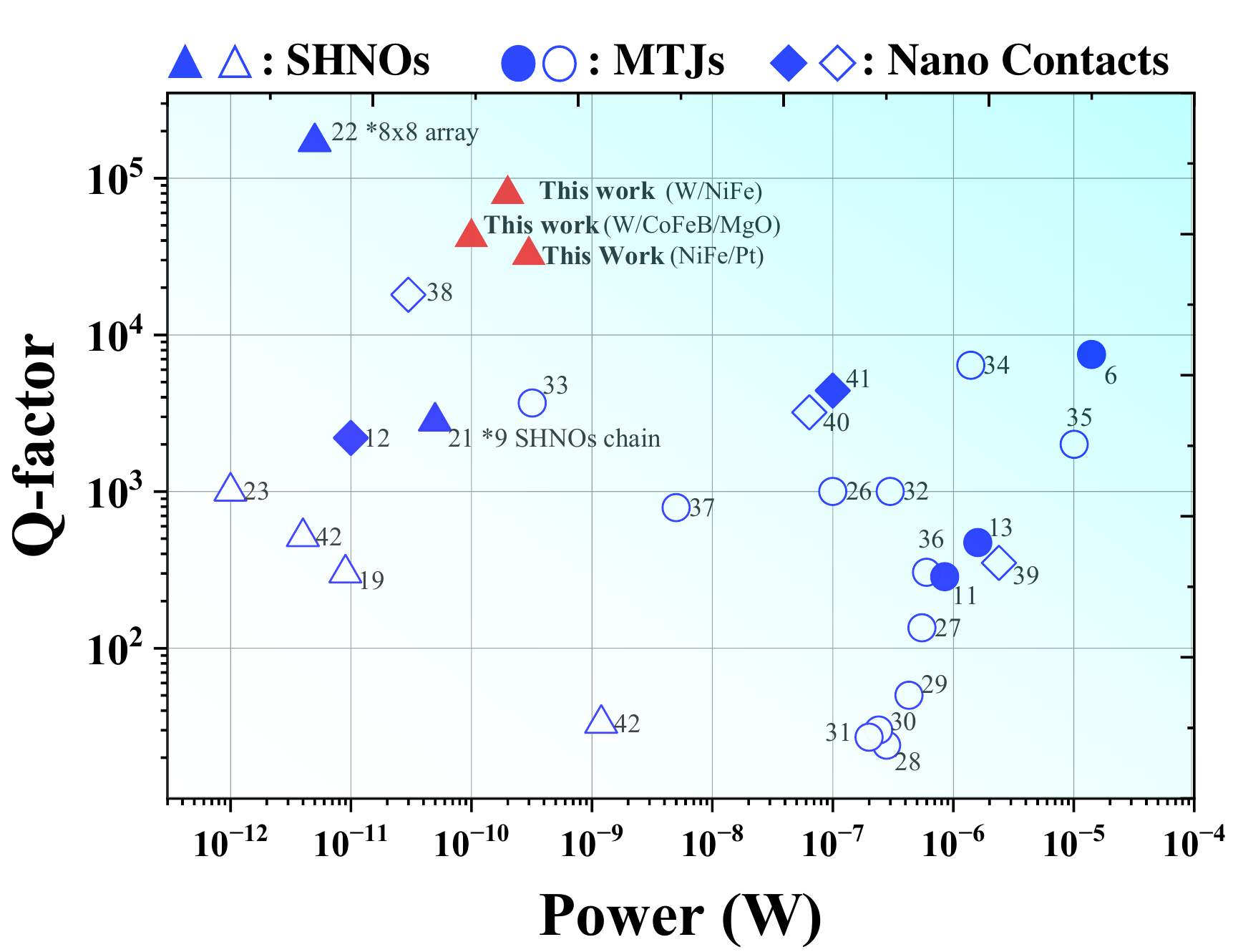}
\caption{\label{fig:Benchmarking} \textbf{Bench-marking of spintronic oscillators}: $Q-$factor \emph{vs.}~ integrated output power of various spintronic oscillators and their synchronized systems (shown with filled symbols). The data comprise the best performance nano-pillar MTJs~\cite{houssameddine2008spin,kubota2013spin,zeng2012high,deac2008bias,zeng2011enhancement,sharma2021electrically,tsunegi2018scaling,costa2017high,seki2014high,lebrun2017mutual}, vortex MTJs~\cite{pribiag2007magnetic,tsunegi2014high,tsunegi2016microwave,dussaux2014large,dussaux2010large}, Nano-contact spin valves~\cite{rippard2004,Maehara2013large,maehara2014high,sani2013mutually,houshang2016spin} and SHNOs~\cite{duan2014nanowire,awad2016natphys,fulara2019spin,zahedinejad2020two,chen2020spin}.}
\end{figure*}

Figure~\ref{fig:parameters}(a) and (b) shows the variation of linewidth and peak power with the number of oscillators for both the W/CoFeB/MgO and W/NiFe system. We observe a 1/$N$ dependence for the spectral linewidth, in agreement with the oscillator synchronization theory and depicts the enhancement of total mode volume. The peak power is found to follow a quadratic ($N^2$) dependence, also consistent with the nonlinear oscillator theory. 

\subsection{Benchmarking the Quality Factor} 
The combination of high auto-oscillation frequencies and low linewidths leads to very high quality factors ($Q=f/\Delta f$) for the synchronized chains. Figure~\ref{fig:parameters}(c) shows $Q$ \emph{vs.}~$N$ for both the W/CoFeB/MgO and W/NiFe systems. We observe $Q>$ 79,000 for 21 mutually synchronized SHNOs in W/NiFe thin films, which is the highest quality factor reported for oscillators in a single chain (which also results in higher output power). This is comparable to the earlier observed $Q$ of 170,000 in two-dimensional NiFe/Pt arrays of 8$\times$8 oscillators. For the W/CoFeB/MgO oscillators, we found a $Q$ of 41,000, which is again the highest of any spintronic oscillator chains operating at frequencies higher than 15 GHz. For comparison with mutually synchronized oscillators based on magnetic tunnel junctions, recent demonstrations~\cite{tsunegi2018scaling} of 8 mutually synchronized MTJs resulted in a $Q-$factor of 7400.

\begin{figure*} [t!]
\centering
\includegraphics[width=14cm]{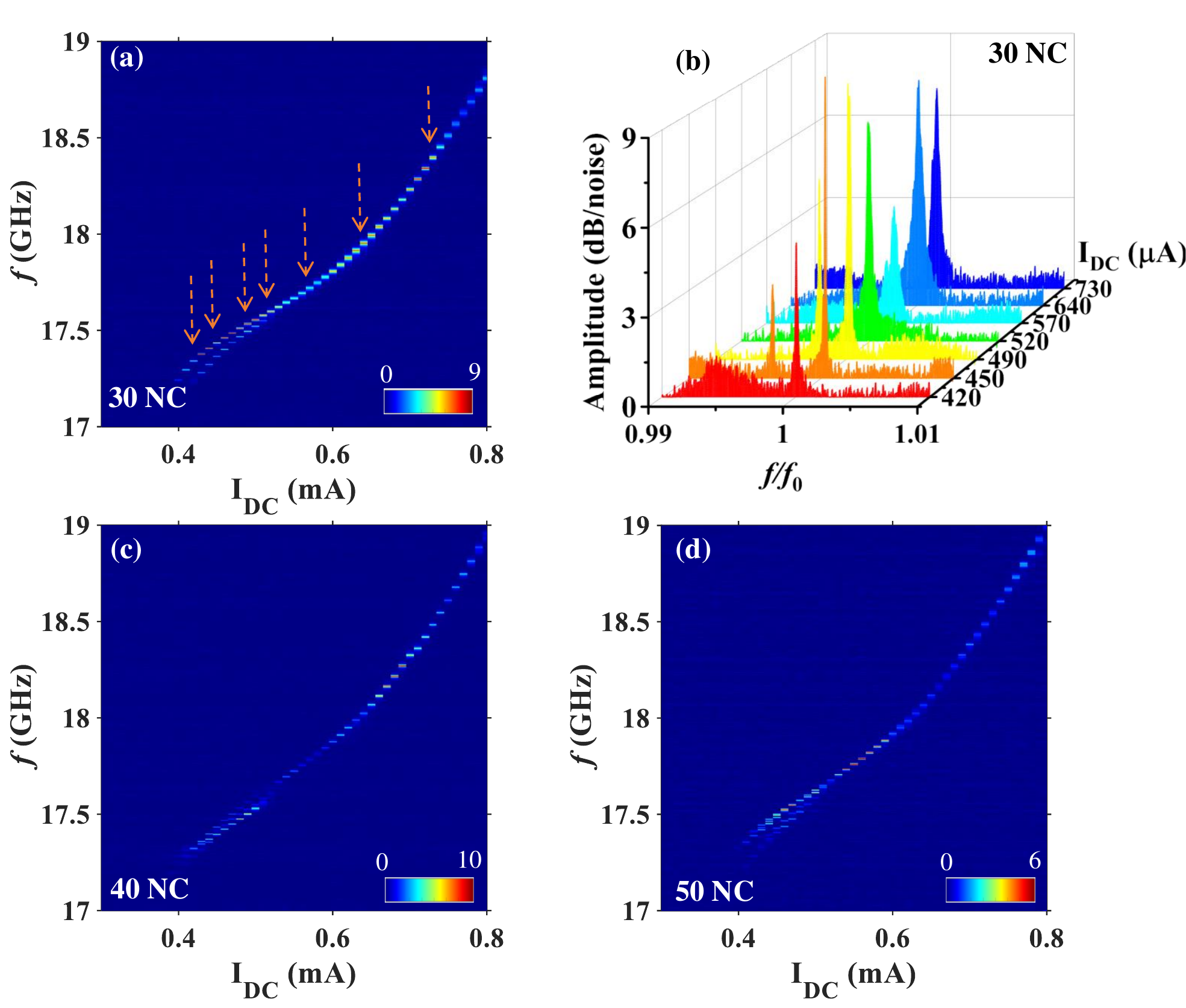}
\caption{\label{fig:beyond20} (a) Power spectral density for synchronized 30 NC SHNOs in a chain. (b) The microwave spectrum for synchronized 30 nano-constrictions in a chain for few current values (I$_{DC}$) marked with arrows in Fig. (a). Power spectral density for mutually synchronized (c) 40 nano-constrictions and (d) 50 nano-constrictions SHNO in a chain for W/CoFeB/MgO thin films.}
\end{figure*}

It is interesting to compare $Q-$factor (and output power) more extensively with the literature. In Fig.~\ref{fig:Benchmarking}, we show results from a large number of previous works, expressed as $Q-$factor \emph{vs.}~output power, for both single (hollow symbols) and synchronized (solid symbols) oscillators. Both here and further on we use the abbreviations 'SHNOs': spin Hall nano-oscillators (shown with triangles), 'MTJs': nano-pillar magnetic tunnel junctions which include vortex oscillators (shown with circular symbols), and 'Nano Contacts': nano-contact spin valve structures (shown with rhombus). One can observe that the vortex oscillators~\cite{pribiag2007magnetic,tsunegi2014high,tsunegi2016microwave,dussaux2010large,dussaux2014large} 
deliver the largest output power (most lie on the right side of the graph). However, vortex oscillators mostly operate at much lower RF frequencies (0.1 GHz to 1.5 GHz) and hence result in lower quality factors. Other MTJs and nano-contact devices have larger frequency tunability but have lower output power and/or higher operational linewidth resulting in poor performance as signal generator. The synchronized MTJs have shown good results and are even employed for power harvesting. Though they also end up with large linewidth and hence low $Q-$factor. The best $Q-$factor found for nano-contact spin valve was about 18000~\cite{rippard2004}, which resulted in poor output power. The recently introduced SHNOs have already shown great promise with their narrow linewidth and high frequency operation, even a single oscillator results $Q-$factor in the range of 2000-4000~\cite{duan2014nanowire,zahedinejad2018cmos,fulara2019spin}, though their output power is extremely low. In our previous work with 2D arrays of 64 oscillators, we observed an enormous $Q-$factor of 179,000, though in a white noise regime (measured at short time scales). In the present work, we have improved our output power and can reach upto 200 pW with sustaining a high $Q-$factor of $>$79000 (for W/NiFe thin films). The NiFe/Pt system shows even higher output power of 300 pW with Q $>$ 32,000. Further improvements in extending synchronization to rectangular arrays and fabricating tunneling magneto-resistance based readouts will bring these oscillators among the best-performing spintronic oscillators with large output power and high $Q-$factor.

\subsection{Synchronization beyond 21 SHNOs}
To investigate synchronization beyond 21 SHNOs, we fabricated 30, 40, and 50 nano-constrictions in series. It is noteworthy that we do observe single-frequency microwave signal generation also in these much longer oscillator chains. However, no further linewidth reduction nor any further increase in the output power were observed. This could be due to an increasing out-of phase synchronization of SHNOs in the longer chains, as illustrated by the spins in Figure 1, if there is a small relative phase shift between individual nearest neighbor SHNOs that does not affect shorter chains but build up to a reduced total output power in the longer chains. Figure~\ref{fig:beyond20} shows the PSD for 30, 40 and 50 nano-constrictions in chain for W/CoFeB/MgO thin films. For 30 nano-constrictions in a chain [Fig.~\ref{fig:beyond20}(a)], we found the lowest linewidth of 800 kHz and a peak power of less than 400 nV$^{2}$/Hz, which is one order of magnitude less than the robustly synchronized 21 SHNOs as shown in Fig.~\ref{fig:parameters}. Figure~\ref{fig:beyond20}(b) shows the transition of partially synchronized microwave emissions into single synchronized mode at different I$_{\rm DC}$. However, as observed in the Fig.~\ref{fig:beyond20}(b) the spectrum at 450 $\mu$A and 490 $\mu$A (partially synchronized states) show lower linewidth and higher amplitude than fully synchronized states at 540 $\mu$A, 640 $\mu$A and 730 $\mu$A current. This clearly shows that the partial synchronization of oscillators results in better spectral parameters than the full synchronization of oscillator chains with more than 30 nano-constrictions. In other words, the longer chains may very well synchronize completely but be better described as weaker synchronization of partially synchronized sub-sections. This lack of robust synchronization for more than 21 nano-constrictions may arise from statistical effects in a larger ensemble, or be due to larger Joule heating in longer chains, and/or result from the increase of an accumulative phase difference in the chains. Figure~\ref{fig:beyond20}(c) and (d) show similar results for 40 and 50 nano-constrictions in a chain, respectively. Though full synchronization in longer chains of more than 21 nano-constrictions is not very robust, it still shows a clear interaction between oscillators (either in-phase or out-of phase), which can be very useful for neuromorphic computing using a large number of spins (nano-constrictions).

\subsection{Perspective: Applications and Outlook}
We have successfully demonstrated the robust mutual synchronization of large number of SHNOs in a single chain. This observation leads to various applications that can be realized using these oscillators. The lower linewidth and the significantly larger output power enable these oscillators for coherent frequency signal generation applications, as well as for wireless communications. Using the phase locked loop method, the microwave oscillations in chains can be further stabilized, generating coherent oscillations that can be directly implemented in many microwave applications~\cite{tamaru2015extremely}. The mutual synchronization of these oscillators in a chain can be explored for bio-inspired computing and beyond~\cite{torrejon2017neuromorphic,zahedinejad2020two,houshang2022prappl} where each oscillator behaves as a neuron. Combined with voltage~\cite{fulara2020giant,kumar2022fabrication,choi2022voltage} and/or memristive~\cite{Zahedinejad2022natmat} control of synchronization (synaptic weights), these large chains can be used to locally or globally control the coupling between oscillators (neurons). The present work also serves as a stepping stone in the direction towards further scaling mutual synchronization to much larger square or rectangular arrays well beyond the previously demonstrated 64 oscillators. As these oscillator arrays can be fabricated in a tiny area, it enables the possibility to scale/miniaturize the neuromorphic networks based on these oscillators. The large frequency tunability with current also makes these oscillators ideal for ultrafast sweeping spectrum analysis, where neither vortex oscillators nor uniform MTJs can so far offer a wide resolution bandwidth~\cite{litvinenko2020ultrafast,litvinenko2022ultrafast}. With recent demonstration of energy-efficient spin Hall materials~\cite{behera2022energy} and the reduction of constrictions size~\cite{durrenfeld2017nanoscale,awad2020width}, the required threshold current can be significantly reduced, operating these devices with ultra-low power.

\section{Conclusion}
In summary, we show that robust in-phase mutual synchronization of nano-constriction based SHNOs can be achieved in very long chains. The long-range synchronization not only shows an enhanced output power but also an improved linewidth of as low as 130 kHz for W/NiFe based heterostructures. The low current and low field operation of these oscillators along with their large frequency tunability, with both current and magnetic field, make them ideal for various spintronic applications such as neuromorphic computing. In the longest chains, mutual synchronization is less effective in improving the microwave signal properties with evidence of partial and/or increasingly out-of-phase synchronization. These results not only enhance our understanding of the mutual synchronization of these oscillators but also paves the way towards making larger networks of these oscillators for neuromorphic computing applications. 

\section*{Methods}
\textbf{\textit{Sample fabrication:}} We utilize the well studied W(5 nm)/CoFeB(1.4 nm)/MgO (2 nm) and W(5 nm)/NiFe(3 nm) heterostructures for the fabrication of microwave nano-constriction SHNOs used in the experiments. The NM/FM structures were deposited using magnetron sputtering on a high resistance intrinsic Si substrate ($\rho >$ 10,000 $\mu$ohm-cm) at room temperature. The sample stacks were capped with 4 nm Al$_{3}$O$_{3}$ thin films. The growth of thin films was carried out using AJA Orion 8 sputtering system with a base pressure of 3$\times$10$^{-8}$ Torr. The samples were then coated with 40 nm of hydrogen silsesquioxane (HSQ) negative tone electron beam resist. The SHNO chains used in the experiments were then fabricated using a combination of e-beam lithography (Raith EBPG 5200) and Ar-ion etching: more details can be found here~\cite{kumar2022fabrication}. The top contact pads are fabricated using laser writer based direct lithography followed by deposition of Cu(800 nm)/Pt(20 nm) thin films for Ground-Signal-Ground co-planar waveguide. In this experiment, we utilize the SHNOs with 150 nm nano-constriction width and 200 nm separation between SHNOs in a chain.

\textbf{\textit{Experimental Set-up:}} All the measurements were performed at room temperature. The microwave measurements were performed using a custom-built probe station utilizing GSG probes manufactured by GGB Industries. Figure 2(c) shows the schematic representation of the measurement setup. All measurements are carried out at a fixed IP angle and an OOP rotatable sample stage between the electromagnet poles at room temperature. Different OOP angles are used to generate positive non-linearity in the system. To excite microwave emission, a positive DC current I$_{DC}$ was applied to the devices through the inductive port of a bias-tee, while the microwave signal was detected using high frequency port. The resulting power spectral density (PSD) of the auto-oscillating signal (after amplification using low-noise amplifier) is captured using Rohde and Schwarz (10 Hz-- 40 GHz) spectrum analyzer. To calculate the PSD spectrum shown in Figure 3 and 7, we have taken into account the impedance mismatch and the losses from the RF component and corrected to the low noise amplifier gain. 

\begin{acknowledgments}
This work was partially supported by the Horizon 2020 research and innovation program (ERC Advanced Grant No. 835068 “TOPSPIN”). This work was also partially supported by the Swedish Research Council (VR) and the Knut and Alice Wallenberg Foundation.
\end{acknowledgments}

\bibliography{main}
\end{document}